\documentclass[
  journal=pasa,
  manuscript=research-paper, %% or "review"
  year=2020,
  volume=37,
]{cup-journal}

\usepackage{microtype,siunitx,booktabs}
\usepackage{amssymb}
\title{Very long baseline interferometry observations of the high-redshift blazar candidate J0141$-$5427}

\author{K.\,\'E.~Gab\'anyi}
\affiliation{Department of Astronomy, Institute of Geography and Earth Sciences, ELTE E\"otv\"os Lor\'and University, P\'azm\'any P\'eter s\'et\'any 1/A, H-1117 Budapest, Hungary}
\alsoaffiliation{Konkoly Observatory, ELKH Research Centre for Astronomy and Earth Sciences, Konkoly Thege Mikl\'os \'ut 15-17, H-1121 Budapest, Hungary}
\alsoaffiliation{ELKH-ELTE Extragalactic Astrophysics Research Group, ELTE E\"otv\"os Lor\'and University, P\'azm\'any P\'eter s\'et\'any 1/A, H-1117 Budapest, Hungary}
\email[K. \'E. Gab\'anyi]{k.gabanyi@astro.elte.hu}

\author{S.~Belladitta}
\affiliation{INAF, Osservatorio Astronomico di Brera, Via Brera 28, 20121 Milano, Italy}
\alsoaffiliation{DiSAT, Universit\`a degli Studi dell'Insubria, Via Valleggio 11, 22100 Como, Italy}
\alsoaffiliation{Max-Planck-Institute for Astronomy, K\"onigstuhl 17, 69117 Heidelberg, Germany}

\author{S.~Frey}
\affiliation{Konkoly Observatory, ELKH Research Centre for Astronomy and Earth Sciences, Konkoly Thege Mikl\'os \'ut 15-17, H-1121 Budapest, Hungary}
\alsoaffiliation{CSFK, MTA Centre of Excellence, Konkoly Thege Miklós \'ut 15-17, H-1121 Budapest, Hungary}
\alsoaffiliation{Institute of Physics, ELTE E\"otv\"os Lor\'and University, P\'azm\'any P\'eter s\'et\'any 1/A, H-1117 Budapest, Hungary}

\author{G.~Orosz}
\affiliation{Joint Institute for VLBI ERIC, Oude Hoogeveensedijk 4, 7991 PD Dwingeloo, The Netherlands}
\alsoaffiliation{School of Natural Sciences, University of Tasmania, Private Bag 37, Hobart, Tasmania 7001, Australia}

\author{L.\,I.~Gurvits}
\affiliation{Joint Institute for VLBI ERIC, Oude Hoogeveensedijk 4, 7991 PD Dwingeloo, The Netherlands}
\alsoaffiliation{Faculty of Aerospace Engineering, Delft University of Technology, Kluyverweg 1, 2629 HS, Delft, The Netherlands}
\alsoaffiliation{CSIRO Astronomy and Space Science, PO Box 76, Epping, NSW 1710, Australia}

\author{K.~Rozgonyi}
\affiliation{University Observatory, Faculty of Physics, Ludwig-Maximilians-Universit\"at, Scheinerstr. 1, 81679 Munich, Germany}
\alsoaffiliation{International Centre for Radio Astronomy Research,
The University of Western Australia, Crawley, WA 6009, Australia}
\alsoaffiliation{Australian Research Council, Centre of Excellence for All-Sky Astrophysics in 3 Dimensions (ASTRO 3D), Canberra, ACT 2611, Australia}

\author{T.~An}
\affiliation{Shanghai Astronomical Observatory, CAS, Nandan Road 80, Shanghai 200030, PR China}
\alsoaffiliation{Peng Cheng Laboratory, Shenzhen 518066, PR China}

\author{H.~Cao}
\affiliation{School of Physics and Electronic Information, Huanggang Normal University, 146 Xingang 2nd Road, Huanggang, Hubei 438000, PR China}

\author{Z.~Paragi}
\affiliation{Joint Institute for VLBI ERIC, Oude Hoogeveensedijk 4, 7991 PD Dwingeloo, The Netherlands}

\author{K.~Perger}
\affiliation{Konkoly Observatory, ELKH Research Centre for Astronomy and Earth Sciences, Konkoly Thege Mikl\'os \'ut 15-17, H-1121 Budapest, Hungary}
\alsoaffiliation{CSFK, MTA Centre of Excellence, Konkoly Thege Miklós \'ut 15-17, H-1121 Budapest, Hungary}

\doi{}

\received {dd Mmm 2022}
\revised  {dd Mmm YYYY}
\accepted {dd Mmm YYYY}
\published{dd Mmm YYYY}

 %% First letter not capped
\begin{document}

\begin{abstract}
Active galactic nuclei (AGN) have been observed as far as redshift $z\sim7$. They are crucial in investigating the early Universe as well as the growth of supermassive black holes at their centres. Radio-loud AGN with their jets seen at a small viewing angle are called blazars and show relativistic boosting of their emission. Thus, their apparently brighter jets are easier to detect in the high-redshift Universe.
DES\,J014132.4$-$542749.9 is a radio-luminous but X-ray weak blazar candidate at $z=5$. We conducted high-resolution radio interferometric observations of this source with the Australian Long Baseline Array at $1.7$ and $8.5$\,GHz. A single, compact radio emitting feature was detected at both frequencies with a flat radio spectrum. We derived the milliarcsecond-level accurate position of the object. The frequency dependence of its brightness temperature is similar to that of blazar sources observed at lower redshifts. Based on our observations, we can confirm its blazar nature. We compared its radio properties with those of two other similarly X-ray-weak and radio-bright AGN, and found that they show very different relativistic boosting characteristics.
\end{abstract}

\section{Introduction}
\label{sec:int}

Active galactic nuclei (AGN) are the most luminous persistent astronomical objects, and they are invaluable probes for investigating the high-redshift Universe. Roughly ten per cent of AGN are radio-loud, jetted sources \citep[e.g.,][]{rl_fraction}. In them, the radio emission originates from the synchrotron emission of the jets. When the jets are seen at a small angles to the line of sight, e.g., $\lesssim10^\circ$ \citep{urry_padovani}, relativistic beaming causes significant flux density enhancement of the advancing jet. Thus, these beamed sources called blazars can be preferentially detected even at high redshifts $(z\gtrsim4)$ in radio bands. 

Blazars can be identified using high-resolution very long baseline interferometry (VLBI) radio observations. They are characterized by a bright feature that is compact at milliarcsec (mas) scale, the jet base, which usually has a flat radio spectrum at GHz frequencies \citep[e.g.,][]{hovatta_spectralshape}. The apparent brightness temperature of this dominant component exceeds the equipartition limit, $T_\mathrm{B}^\mathrm{eq} \approx 5 \times 10^{10}$\,K \citep{equipartition}, and sometimes even the inverse Compton limit \citep[$\sim 10^{12}$\,K,][]{inverseCompton}, indicating the potential prevalence of relativistic beaming. Also, apparent superluminal motion of components can often be observed in blazar jets. However, in the case of high-redshift sources, the steep-spectrum jet components are harder to detect, because the observed frequencies correspond to $(1+z)$ times higher emitted frequencies in the source's rest frame, thus the extended regions of jets are often undetectably faint \citep{Gurvits_jet}. 

Blazars can also be classified via their broad-band spectral energy distribution (SED) featuring non-thermal emission over the electromagnetic spectrum and exhibiting relativistic beaming effect \citep[e.g.,][]{roma_bzcat}.

\citet{Belladitta_2019} reported the discovery of DES J014132.4$-$542749.9 (hereafter J0141$-$5427), a radio-bright but X-ray-weak AGN at $z=5.00\pm0.01$. The source, according to archival data and newly obtained X-ray observations of the authors, is an order of magnitude fainter in X-rays than other blazars with similar radio luminosities.  
\cite{Belladitta_2019} showed that the SED of J0141$-$5427 can be best described with a relativistically beamed blazar SED if a very high magnetic field strength of $\sim 9$\,G is assumed.  

We initiated VLBI observations of J0141$-$5427 with the Australian Long Baseline Array (LBA) at $1.7$ and $8.5$\,GHz to ascertain its blazar nature.

Hereafter we use the flat $\Lambda$CDM cosmological model with parameters of $H_0=70\, \mathrm{km\,s}^{-1}\mathrm{\,Mpc}^{-1}$, $\Omega_\mathrm{m}=0.27$, and $\Omega_\Lambda=0.73$. At the redshift of J0141$-$5427, $1$\,mas angular size corresponds to a projected linear length of $\sim6.5$\,pc, and the luminosity distance of the object is $D_\textrm{L}=48273.2$\,Mpc \citep{Cosmocalc}.

%--------------------------------------------------------------------
\section{Observations and data reduction} \label{sec:obs}

Observations of J0141$-$5427 with the LBA were conducted in 2020, under the project code v591 (PI: K. \'E. Gab\'anyi) in phase-referenced mode \citep{p-ref}. In this observing mode, the pointing directions of the telescopes change regularly between the target source and a nearby phase-reference calibrator within the atmospheric coherence time permitted by the radio propagation media. The delay, and delay rate solutions can be then transferred (interpolated) from the calibrator to the target source. The nodding cycles in both the $1.7$ and $8.5$\,GHz observations were $5$\,min long, with $3.5$\,min spent on the target and $1.5$\,min on the calibrator. The phase-reference calibrator was ICRF~J015649.7$-$543948 in both experiments. Additional calibrator sources were also observed to facilitate amplitude calibration, and to monitor the stability of the array. 

The $1.7$~GHz observation took place on 2020 June 26 and 27, the participating antennas were Ceduna (CD), Hobart (HO), Mopra (MP), Parkes (PA), the tied array of the Australia Telescope Compact Array (ATCA) in Australia, and Hartebeesthoek (HH) in South Africa. The observation lasted for $10$\,h, the on-target time was $4.5$\,h. The $8.5$~GHz observation took place on 2020 July 14, with the following participating antennas: CD, HO, MP, PA, Katherine (KE), Yarragadee (YG), the tied array of ATCA in Australia,  the $12$-m Warkworth antenna (WW) in New Zealand, and HH in South Africa. The observation lasted for $10.25$\,h, with an on-target time of $4.9$\,h. In both observations, the total bandwidth of $128$\,MHz was divided into $8$ intermediate frequency bands (IF) of $32$ channels each. The correlator integration time was set to $2$\,s. The correlation was done at the Pawsey Supercomputing Centre in Perth, on a DiFX software correlator \citep{Deller2011}. The longest baselines of the arrays (providing the finest angular resolution) were those to HH. At $8.5$\,GHz, HH could only participate in the last $18$\,min of the observation.

Data reduction was done using the National Radio Astronomy Observatory (NRAO) Astronomical Image Processing System (\textsc{aips}, \citealt{aips}) following standard procedures of ionospheric and parallactic angle corrections, manual phase calibrations and fringe-fitting of the calibrator sources, and following the LBA guide on amplitude calibration\footnote{\url{https://www.atnf.csiro.au/vlbi/dokuwiki/doku.php/lbaops/lbacalibrationnotes} (accessed 2022.09.06)}. The necessary files for amplitude calibration were created from the system temperature measurements and gain curves provided by the participating stations or system equivalent flux densities listed in the LBA amplitude calibration user's guide. In the absence of system temperature measurements (at the antennas CD and HO at $1.7$\,GHz and HO, KE, and WW at $8.5$\,GHz), nominal system temperature values were used.

The fringe-fitting was performed for all calibrator sources. Solutions were found for $\sim 86$\% and $\gtrsim 98$\% of the data at $1.7$\,GHz and $8.5$\,GHz, respectively.

At the AT, MP, and PA antennas, wider filters were used, resulting in clearly lower amplitude values in the channel-averaged data at those IFs corresponding to the edges of the bands (IFs 1, 4, 5, 8) compared to the ones at the middle (IFs 2, 3, 6, 7). Therefore, the edge IFs were scaled up by a constant factor of $1.169$ to bring them closer to the values measured in the middle of the band at these antennas in the case of $1.7$-GHz observation, before channel averaging. At the $8.5$-GHz observation, instead of such scaling, we flagged the first $10$ channels for IFs 1 and 5 and the last $10$ channels for IFs 4 and 8, for the three antennas using wider filters (AT, MP, and PA). 

After the fringe-fitting performed on the calibrator sources, and the application of the above described amplitude scaling for the $1.7$-GHz amplitudes, the channel-averaged data of the calibrator sources were imported into the Caltech \textsc{Difmap} package \citep{difmap} for hybrid mapping. The hybrid-mapping procedure involves subsequent steps of \textsc{clean}ing \citep{clean} and phase self-calibration of the data. As the last step, amplitude self-calibration was done. The gain correction factors obtained for different calibrator sources were in good agreement for the same antennas and IFs. The flux density values of the phase-reference calibrator obtained this way at both frequencies were in good agreement with the ones measured by the ATCA closest in time and at similar frequencies according to the ATCA Calibrator Database\footnote{\url{https://www.narrabri.atnf.csiro.au/calibrators/calibrator_database.html} (accessed 2022.09.06)} with the VLBI-measured flux densities $\sim15$\,\% and $\sim3$\,\% lower than the ones measured by ATCA at $1.7$\,GHz and at $8.5$\,GHz, respectively. The difference is most probably caused by resolution effect, the LBA observations resolved out the large-scale emission detected by ATCA. Thus, we accepted the gain correction factors obtained in \textsc{Difmap} for the phase-reference calibrator, and we adjusted the antenna gains accordingly in \textsc{aips} to further improve the amplitude calibration. 

To improve the delay and rate solutions, the phase-reference calibrator was fringe-fitted again using the \textsc{clean} component model of its brightness distribution derived from the hybrid mapping, to take the source structure into account. The obtained solutions were applied to the phase-reference calibrator as well as to the target source, and subsequently both were imaged in \textsc{Difmap}. 

In the case of the phase-reference calibrator, the amplitude self-calibration performed after this second hybrid mapping showed that the gain correction factors were mostly $\lesssim 10$\% for the $1.7$-GHz data and mostly $\lesssim 5$\% for the $8.5$-GHz data, except for single IFs of AT and CD, and a few discrepant IFs of YG. Additionally, it seemed that amplitude self-calibration of HH was not constrained at $1.7$-GHz, and it could not correct the amplitudes. Thus, we conservatively assume the amplitude calibration of these LBA data is reliable at $10$\% level.

Due to an unfortunate typing mistake made by the PI at the time of scheduling, the observations and subsequent correlations were done at a target source position with $4^{\prime\prime}$ offset in declination from the previously known position. A significant offset from the phase centre may cause reduction of the peak intensity and distortion of the obtained image through bandwidth smearing and time-average smearing effects \citep{smearing_1999}. 

The bandwidth smearing effect would have been substantial (intensity reduction of a point source by $\sim 80-90$\,\%) if the data were averaged over all the channels within an IF \citep{smearing_1999,smearing_wrobel}. Therefore, the hybrid mapping of the target source was performed on the unaveraged data. We disregarded the first and last $5$ channels of all $8$ IFs to account for bandpass effects. 

At both frequencies, time averaging was done for $2$\,s at the correlator. However, because of the different resolutions, time-average smearing affects the two data sets differently. At $1.7$\,GHz, this effect is negligible, the peak intensity reduction of a point source is less than $1$\% at $4^{\prime\prime}$ from the pointing centre. At $8.5$\,GHz, if calculated for the highest achievable resolution obtained with the longest baseline, between HH and YG, time-average smearing would cause an average peak intensity reduction of a point source by $15$\%. Excluding the baselines to HH, the average amplitude reduction of a point source is $\sim 5$\% at $4^{\prime\prime}$ distance. Since HH could only participate in the last $18$\,min of the $8.5$\,GHz observation, we excluded the data on the baselines to HH. Therefore the effects of the unintentional pointing offset introduced in the target source position could be mitigated satisfactorily.

The target source J0141$-$5427 turned out to be bright enough for attempting a direct fringe-fitting. Before that, the visibility data set was shifted by $4^{\prime\prime}$ in declination direction to its a priori known correct position using the task \textsc{clcor} in \textsc{aips}. At $1.7$\,GHz, fringes with a signal-to-noise level exceeding $6\sigma$ were found for $69$\% of data, including the longest baselines to HH. We continued imaging both the fringe-fitted and the phase-referenced $1.7$-GHz data of the target, and the results were in good agreement. The peak intensity was less by $\sim 10\textrm{\,mJy\,beam}^{-1}$ ($\sim 13$\,\%) in the phase-referenced image compared to the one obtained after fringe-fitting the data due to the coherence loss \citep{coherence-loss}. At $8.5$\,GHz, at the same signal-to-noise level, fringes were found for only $24$\% of data, and no fringes were found on the baselines to HH. Therefore, we did not use the fringe-fitted data of the target for the higher frequency observation.

At both frequencies, phase self-calibration and amplitude self-calibration were performed with subsequently shorter time intervals during the hybrid mapping of J0141$-$5427. However, only the best-behaving, least noisy antennas were used in the self-calibration processes. Thus, HH and CD were kept fixed for the $1.7$-GHz observation. In the case of the $8.5$-GHz observation, originally all antennas were used in phase self-calibration (except for HH which was not used in the hybrid mapping), but for the shortest time intervals, and in the amplitude self-calibration, only ATCA, MP, CD, and PA were included, while the gains of the remaining antennas were kept fixed.

\section{Results}
\label{sec:res}

\begin{figure}
\centering
\includegraphics[width=\textwidth, bb=30 165 570 700, clip]{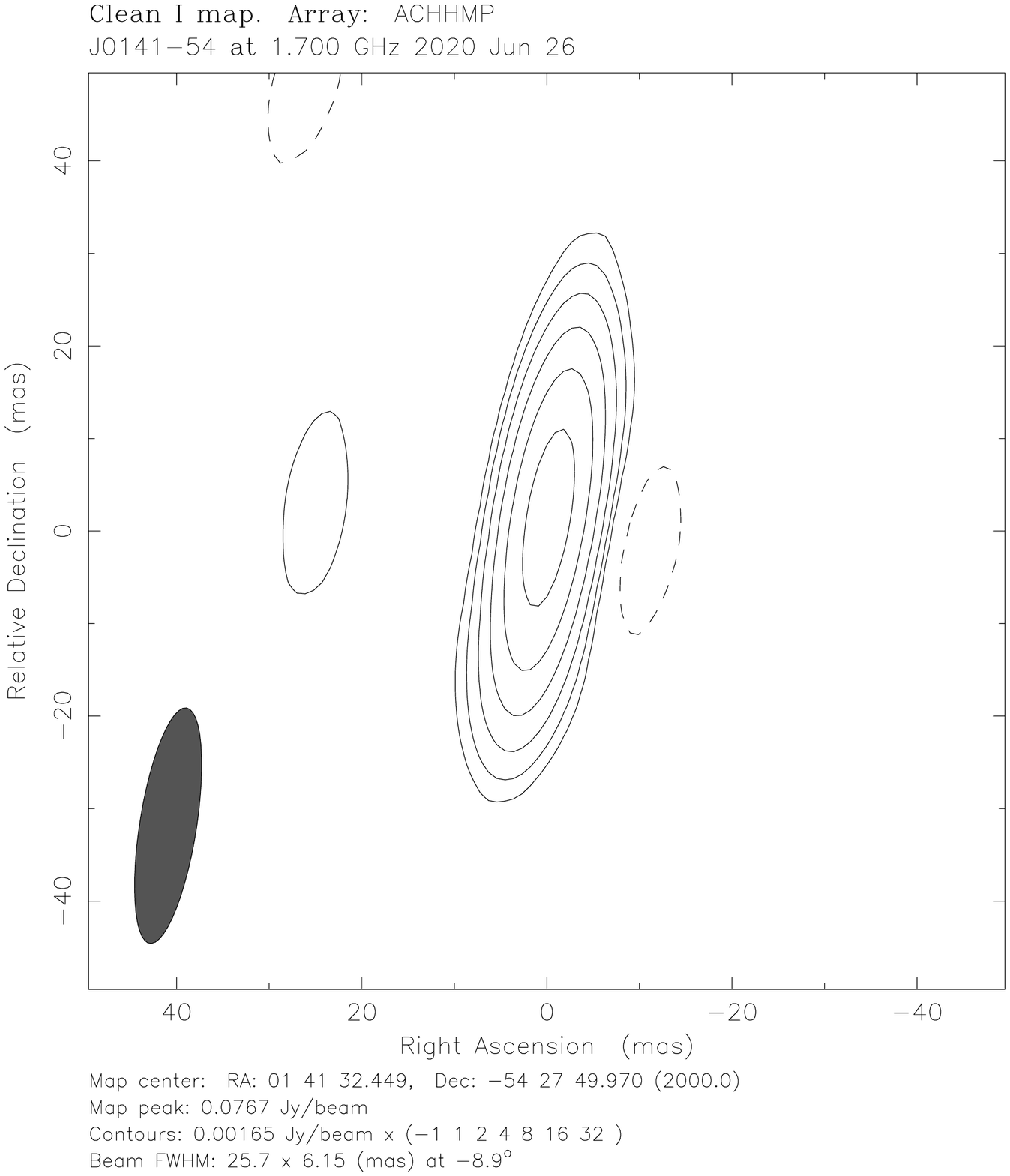}
\caption{$1.7$-GHz naturally-weighted LBA map of the fringe-fitted data of J0141$-$5427. The peak intensity is $76.7\mathrm{\,mJy\,beam}^{-1}$. The lowest contours are at $\pm1.7\mathrm{\,mJy\,beam}^{-1}$, corresponding to $4\sigma$ image noise level. Further positive contours increase by a factor of $2$. The elliptical Gaussian restoring beam size is $25.7\mathrm{\,mas}\times6.2\mathrm{\,mas}$ at a major axis position angle of $-8.9^\circ$, and it is shown in the lower left corner of the image.}
\label{fig:LBA_L}
\end{figure}

\begin{figure}
\centering
\includegraphics[width=\textwidth, bb=30 130 585 680, clip]{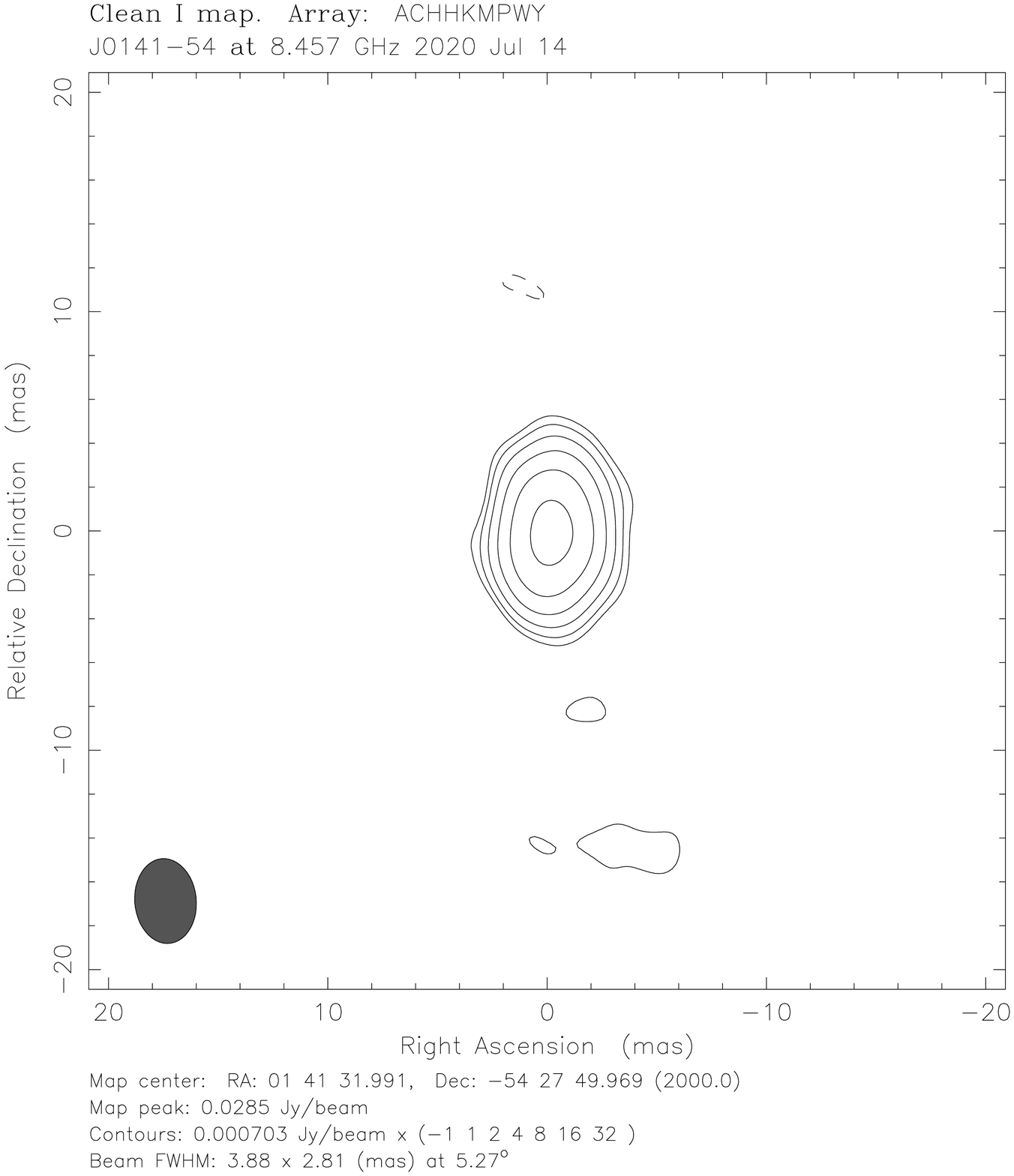}
\caption{$8.5$-GHz naturally-weighted phase-referenced LBA map of J0141$-$5427. The peak intensity is $28.5\mathrm{\,mJy\,beam}^{-1}$. The lowest contours are at $\pm0.7\mathrm{\,mJy\,beam}^{-1}$, corresponding to $4\sigma$ image noise level. Further positive contours increase by a factor of $2$. The elliptical Gaussian restoring beam size is $3.9\mathrm{\,mas}\times2.8\mathrm{\,mas}$ at a major axis position angle of $5.3^\circ$, and it is shown in the lower left corner of the image.}
\label{fig:LBA_X}
\end{figure}

At both frequencies, a single radio-emitting feature was detected (Figs. \ref{fig:LBA_L} and \ref{fig:LBA_X}).

We derived the coordinates of the brightest pixel at both frequencies using the \textsc{aips} verb \textsc{maxfit}. At $8.5$\,GHz, the right ascension and declination are $\mathrm{RA} = 1^\textrm{h} 41^\textrm{m} 32.44937^\textrm{s}$ and $\mathrm{Dec} = -54^\circ 27^\prime 49.9705^{\prime\prime}$, respectively. We estimate that these coordinates are accurate within $0.8$\,mas. The most dominant sources of the uncertainty are the positional accuracy of the phase reference calibrator ($0.37$\,mas in right ascension and $0.33$\,mas in declination direction, according to the most recent version of the Radio Fundamental Catalog\footnote{rfc\_2022b, \url{http://astrogeo.org/sol/rfc/rfc_2022b/rfc_2022b_cat.txt} (accessed 2022.09.06)}) and the astrometric errors strongly depending on the phase-reference calibrator--target angular separation. For the latter, we conservatively assumed the value derived for observations taken at $5$\,GHz by \cite{chatterjee2004}. The coordinates derived from the phase-referenced $1.7$-GHz observation agree with the $8.5$-GHz values within the uncertainties. Additionally, they agree within the uncertainty with the optical position provided in the Dark Energy Survey 2nd data release \citep{des}. These newly derived radio coordinates of J0141$-$5427 are much more accurate than those previously obtained from lower-resolution radio observations, e.g. the AT20 survey with $\sim 1''$ positional accuracy \citep{at20_murphy}.

To quantitatively describe the brightness distribution of the source, we fitted the visibility data with Gaussian model components. At $1.7$\,GHz, a single circular Gaussian component with a flux density of $(80.3\pm8.4)$\,mJy and a full-width at half-maximum (FWHM) size of $\sim 2.2$\,mas can adequately describe the data. However, according to \cite{lister2021}, the typical uncertainty of a single isolated Gaussian brightness distribution component diameter is $20$\% of the restoring beam FWHM size. As such, the size of the component is not well-constrained. The highly elongated restoring beam of the $1.7$-GHz experiment, major axis $25.7$\,mas in roughly north--south direction and minor axis $6.2$\,mas in the perpendicular direction, would result in an asymmetric source size uncertainty in the two perpendicular orientations. In the finer resolution east--west direction, the FWHM size of the emitting feature is $(2.2\pm1.5)$\,mas, while it is not constrained in the perpendicular direction, $(2.2 \pm 5.1)$\,mas. Nevertheless, the compactness of the radio emission is further supported by the high percentage of fringe solutions found on the longest baselines to HH.

At $8.5$\,GHz, an elliptical Gaussian component with a flux density of $(40.8 \pm 4.1)$\,mJy, a major and a minor axis FWHM sizes of $(3.1\pm 0.8)$\,mas and $(1.4\pm 0.6)$\,mas, respectively, and a major axis position angle of $-13.7^\circ$ was needed to fit the data\footnote{Position angles are measured from north through east.}. The $8.5$-GHz observation is somewhat affected by time smearing effect as described in Section~\ref{sec:obs}. While the peak intensity reduction of a point source may not be significant, time-average smearing can cause distortion of the image. Therefore, we also analysed the data set by excluding the longer baselines where the smearing effect is expected to be more pronounced. We only retained the antennas of MP, PA, ATCA, HO, and CD. We obtained the same parameters within the errors for the fitted Gaussian brightness distribution model, suggesting that the modeling results are robust. 

Assuming the same amount of coherence loss we seen at $1.7$\,GHz ($\sim 15$\,\%), the flux density of the detected feature is $(46.9\pm4.7)$\,mJy at $8.5$\,GHz.

\section{Discussion}
\label{sec:dis}

\subsection{Brightness temperature} \label{sec:tb}

The brightness temperature of the source in the rest-frame of the source can be calculated with the following equation \citep[e.g.,][]{tb_veres,hovatta_spectralshape}:
\begin{equation}
    T_\textrm{b}=1.22 \times 10^{12} \frac{S}{\theta_\textrm{maj}\theta_\textrm{min}\nu_\textrm{o}^2}(1+z),
\end{equation}
where $S$ is the flux density in units of Jy, $\nu_\textrm{o}$ is the observing frequency in unit of GHz, and $\theta_\textrm{maj}$ and $\theta_\textrm{min}$ are the major and minor axes (FWHM) of the Gaussian radio-emitting feature in units of mas. The brightness temperature of the modeled feature measured at an observing frequency of $8.5$\,GHz is $T_{\textrm{b, }\nu_\textrm{o}=8.5}=(1.1\pm 0.9) \times 10^{9}$\,K. At $1.7$\,GHz observing frequency, due to the poorly constrained component size, the brightness temperature has much larger error, $T_{\textrm{b, }\nu_\textrm{o}=1.7}=(4.2\pm 3.1) \times 10^{10}$\,K. Despite the large uncertainty, $T_{\textrm{b, }\nu_\textrm{o}=1.7}$ exceeds $T_{\textrm{b, }\nu_\textrm{o}=8.5}$, which would contradict the naive expectations of detecting more compact, thus of higher brightness temperature, emitting feature in higher-resolution VLBI observation. 

However, the $8.5$\,GHz and $1.7$\,GHz observing frequencies correspond to $\sim 51.0$ and $\sim 10.2$\,GHz rest-frame frequencies, respectively, at the redshift of the source ($z=5$). \citet{Cheng2020} studied a large sample (more than $800$ objects) of compact, bright radio-loud AGN (mostly blazars), and showed that $T_\textrm{b}$ at $43$\,GHz and at $86$\,GHz rest-frame frequencies are below the values obtained at lower rest-frame frequencies, between $2$\,GHz and $22$\,GHz, due to synchrotron opacity effect. %In the light of this finding, our results for the brightness temperature of J0141$-$5427 indicate that the synchrotron turn-over frequency, where the highest $T_\textrm{b}$ value is expected, is below the rest-frame frequency of $51$\,GHz.

By analysing data from large multi-frequency VLBI surveys, \cite{Cheng2020} found that the frequency-dependence of the core brightness temperature can be well described with a broken power law (up until $240$\,GHz), with the maximum brightness temperature reached at the break frequency of $\sim 6.8$\,GHz. Using their best fit parameters for the shape of the curve, we obtain a brightness temperature value of $(6.4\pm2.0) \times 10^{10}$\,K at the break frequency of $6.8$\,GHz (rest-frame, corresponding to $1.1$\,GHz observing frequency). If, instead, we fit for both the break frequency, $\nu_\textrm{j}$ and the brightness temperature at $\nu_\textrm{j}$, we obtain $(20.5 \pm 8.2) \times 10^{10}$\,K at a rest-frame frequency $\nu_\textrm{j}=(3.6\pm0.4)$\,GHz (corresponding to an observing frequency of $0.6$\,GHz).

The brightness temperature values of J0141$-$5427 obtained at rest-frame frequencies of $51$ and $10.7$\,GHz clearly indicate that the radio emission is related to the activity of the central supermassive black hole in an AGN, and cannot be explained by star formation in the host galaxy \citep{condon_radio}. At face value, they do not exceed the theoretical equipartition limit of $\sim 5 \times 10^{10}$\,K of \cite{equipartition} and the empirically found median intrinsic brightness temperature of blazar sources, $4.1 \times 10^{10}$\,K \citep{homan_tb_new}. Thus, Doppler boosting is not crucially needed to explain the brightness temperature values, however it cannot be ruled out. On the other hand, taking the decrease in brightness temperature at high rest-frame frequencies well above the break frequency into account, the measured brightness temperatures are compatible with those of a slightly Doppler boosted blazar source.

Brightness temperature values significantly below the equipartition limit usually correspond to physical processes in evolved plasma regions and not in the compact regions of blazar jets. However, such low brightness temperatures can also be measured in blazars due to insufficient resolution, when the core and a close jet component cannot be resolved and thus the fitted size is larger, resulting in lower $T_\textrm{b}$ value. \cite{hovatta_spectralshape} studied $190$ blazar jets of the Monitoring Of Jets in Active galactic nuclei with VLBA Experiments \citep[MOJAVE,][]{Mojave2019} survey, and showed that in sources at higher redshifts\footnote{The highest-redshift objects of this sample are at $z\approx 3.3$ \citep{hovatta_spectralshape}.} the derived core parameters are more likely contaminated by a neighbouring jet component due to the lower effective linear resolution.

Interestingly, high brightness temperature values, close to the equipartion limit are rarely observed in other $z>5$ blazars \citep{Coppejans2016,Zhang2022}, with the notable exception of J0906$+$6930 \citep{natcom}.

\subsection{Flux density and spectral index}

Compared to lower-resolution radio observations of J0141$-$5427, there is a significant difference in the recovered flux density. According to the AT20G survey, the source had a flux density of $(70.0\pm4.0)$\,mJy at $8.6$\,GHz, measured between 2004 and 2008 with the ATCA \citep{chhetri2013}. This discrepancy can be due to resolution effect, i.e. the LBA observation resolving out a significant fraction of extended radio emission, and/or source flux density variability in time.

We can derive the spectral index ($\alpha$) of the compact radio-emitting feature between $1.7$\,GHz and $8.5$\,GHz observing frequencies using the flux densities obtained from the Gaussian model fitting to our LBA visibility data. The spectral index is defined as $S\propto\nu^\alpha$. For J0141$-$5427, $\alpha=-0.33\pm0.13$, thus it has a flat radio spectrum. This is a typical spectral index value for the core of jetted AGN \citep[e.g.,][]{hovatta_spectralshape} between $8$ and $15$\,GHz. 

\begin{figure}
\centering
\includegraphics[width=\textwidth, bb=60 40 725 555, clip]{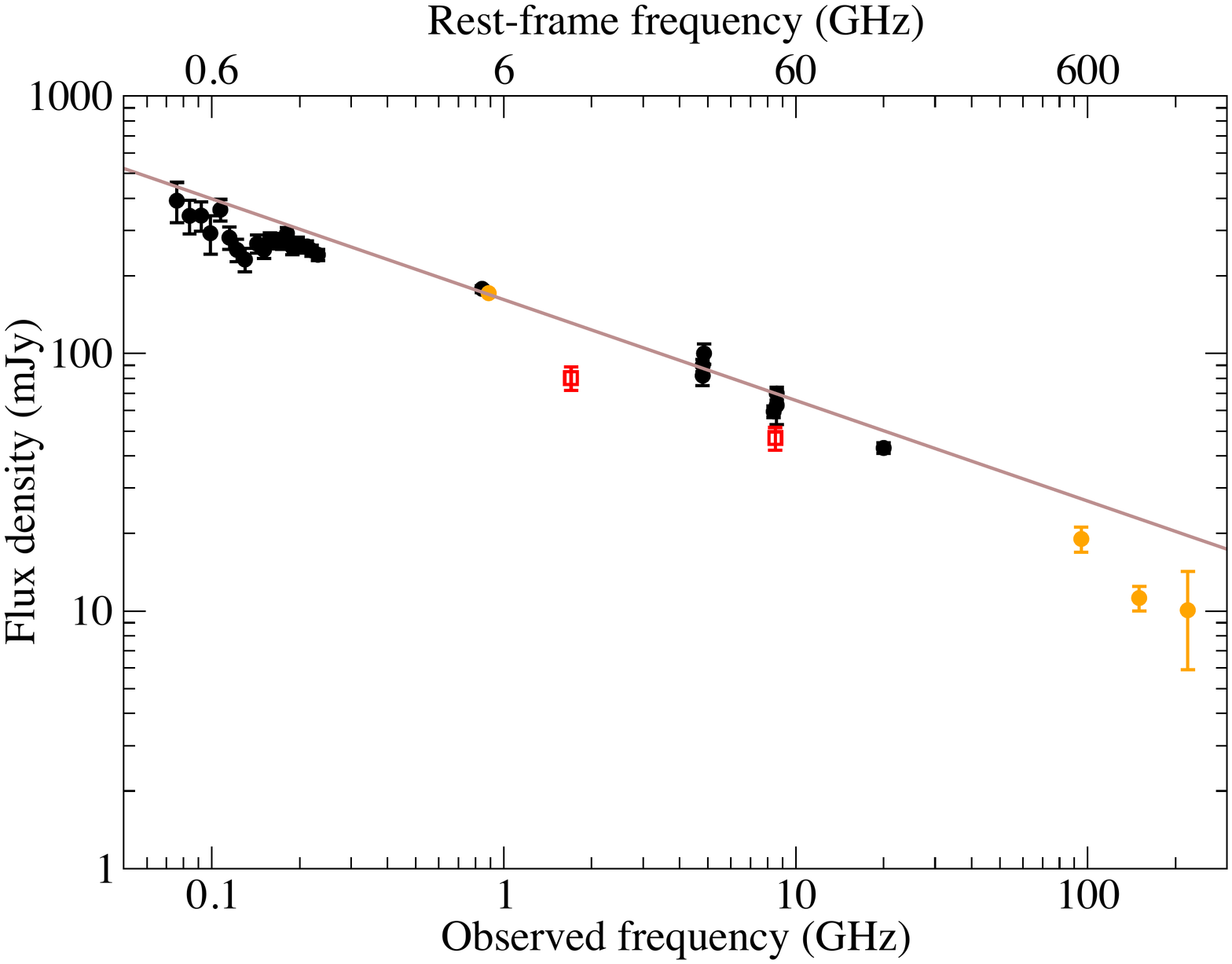}
\caption{Radio spectrum of J0141$-$5427. Black circles are low-resolution archival measurements \citep[for references, see][]{Belladitta_2019}. Orange circles are from the RACS DR1 \citep{racs1,racs2}, and from the SPT-SZ survey \citep{sptsz-cat}. Red squares are our LBA flux densities. The brown line represents a power-law fit to the low-resolution data (black and orange symbols).}
\label{fig:spectrum}
\end{figure}

The source shows a similarly flat radio spectrum between $76$\,MHz and $20$\,GHz in archival low-resolution radio observations as reported by \cite{Belladitta_2019}. Since that publication, the first data release of the Australian Square Kilometre Array (SKA) Pathfinder (ASKAP), the Rapid ASKAP Continuum  Survey \citep[RACS,][]{racs1,racs2} has become public. In addition, we included high-frequency radio flux density measurements obtained with the South Pole Telescope (SPT) within the framework of SPT Sunyaev--Zeldovich survey \citep[SPT-SZ,][]{sptsz-cat}. The most complete radio spectrum of J0141$-$5427 is shown in Fig.~\ref{fig:spectrum}.

\begin{figure}
    \centering
    \includegraphics[width=\textwidth, bb=45 235 550 600, clip]{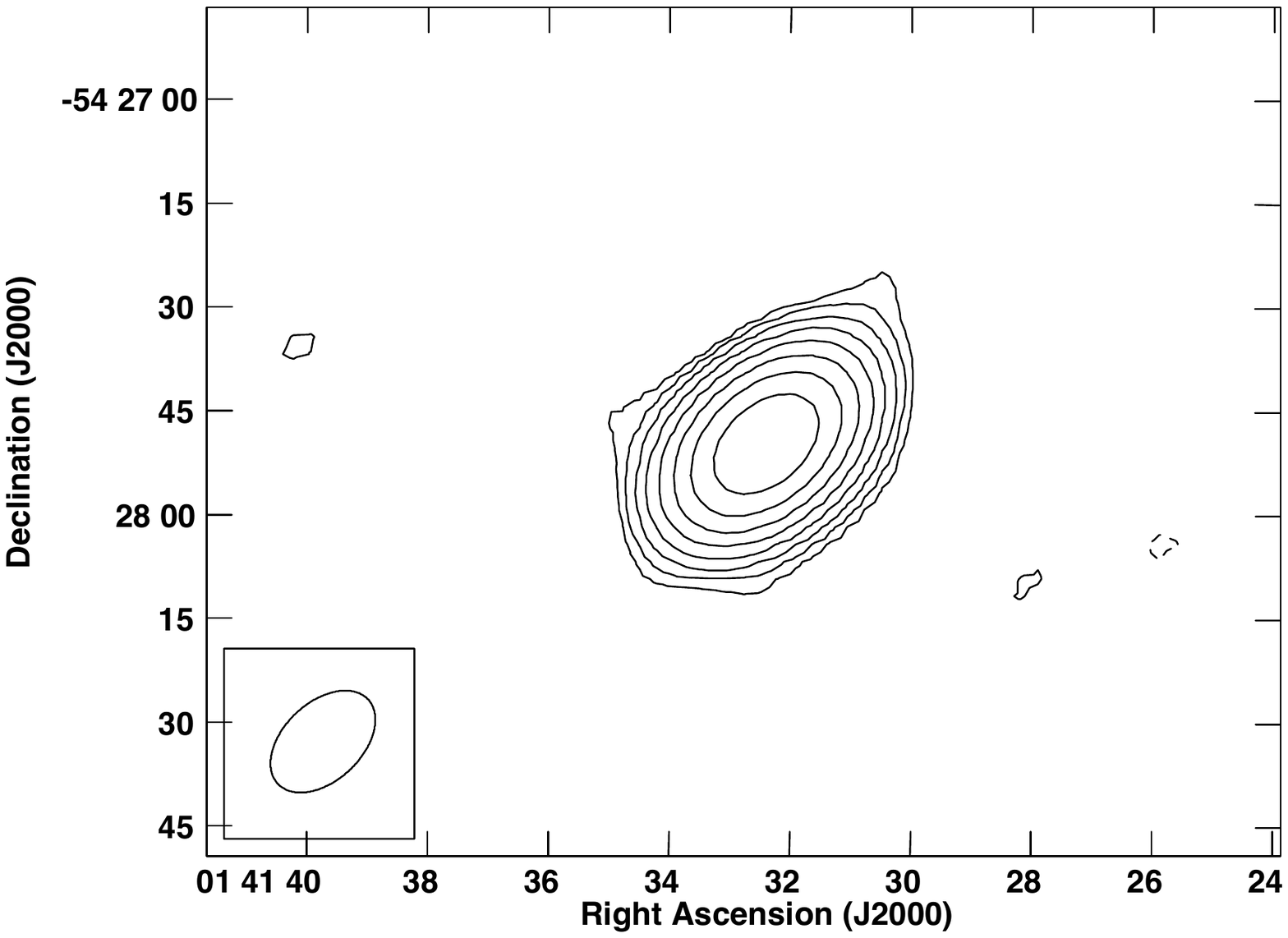}
    \caption{ASKAP image of J0141$-$5427 at $888$\,MHz from RACS \citep{racs1,racs2}. Peak brightness is $162.6\textrm{\,mJy\,beam}^{-1}$. The lowest contours are drawn at $\pm0.68\textrm{\,mJy\,beam}^{-1}$ corresponding to an image noise level of $3\sigma$, further positive contour levels increase by a factor of two. The restoring beam is $17.84^{\prime\prime} \times 11.28^{\prime\prime}$ at a major axis position angle of $-46.7^\circ$, as shown in the lower left corner of the image.}
    \label{fig:racs}
\end{figure}

J0141$-$5427 has been detected in RACS as a single-component source with a flux density of $(174.0 \pm 13.0)$\,mJy at $888$\,MHz (Fig.~\ref{fig:racs}). This value agrees within the errors with the closest-frequency measurements taken at $843$\,MHz by the Sydney University Molonglo Survey \citep[SUMSS,][]{sumss} in 2002.
At higher frequencies, J0141$-$5427 was detected in all three bands of the SPT-SZ, at $95$\,GHz, $150$\,GHz, and $220$\,GHz (however, at the highest frequency only with a signal-to-noise ratio of $2.9$) with flux densities of $S_\textrm{95}=(19.1\pm2.2)$\,mJy, $S_\textrm{150}=(11.2\pm1.2)$\,mJy, and $S_\textrm{220}=(10.1\pm4.2)$\,mJy, respectively. These measurements indicate a possible steepening of the radio spectrum at high frequencies. However, the broad-band radio spectrum is still flat with $\alpha_{0.076}^{220}=-0.39 \pm 0.02$. The observing frequencies of SPT correspond to rest-frame frequencies of $570$\,GHz, $900$\,GHz, and $1320$\,GHz, where the emission from the dust in the host galaxy may have a growing contribution to the measured flux density \citep{Planck2016,Massardi2022}.

There is no sign of spectral turnover of the radio spectrum at observed frequencies around $0.6$\,GHz and $1.0$\,GHz corresponding to the rest-frame turnover values estimated from the brightness temperatures (see Sect. \ref{sec:tb}). There is a hint of spectral flattening at around a few hundred MHz  measured by the GaLactic and Extragalactic All-sky MWA Survey (GLEAM) \citep{Belladitta_2019}, which is followed by a steepening at lower observed frequencies, below $\sim 130$\,MHz. However, this apparent rise of the flux density with decreasing frequency (and thus decreasing angular resolution) could be caused by source confusion; according to \cite{gleam}, confusion is the limiting noise factor at $\lesssim 100$\,MHz in the GLEAM data. The effect of confusion, the target source being blended with its neighbours, has also been seen in lower frequency GLEAM data by An et al. (2022, submitted).

\subsection{Magnetic field strength}

The magnetic field strength of a compact synchrotron self-absorbed source can be estimated if the frequency of the spectral turnover from the optically thick to the optically thin region, and the flux density ($S_\textrm{j}$) and the angular size of the emitting region at the turnover point ($\theta_\textrm{j}$) are known \citep{marscher_turnover},

\begin{equation}
B=10^{-5} b(\alpha) \theta_\textrm{j}^4\nu_\textrm{j}^5 S_\textrm{j}^{-2} \frac{\delta}{1+z},
\end{equation}
where $\delta$ is the relativistic Doppler boosting factor, and $b(\alpha)$ is a numerical factor depending on the spectral index tabulated in \cite{marscher_turnover}.

Using $\nu_\textrm{j}=6.8$\,GHz from \cite{Cheng2020} and assuming that $\alpha$ does not change till the turnover, we can calculate the expected flux density at this (rest-frame) frequency, $S_\textrm{j}=92.1$\,mJy. The size of the emitting region can be derived from the fitted brightness temperature as $\theta_\textrm{j}=3.0$\,mas. Thus, the magnetic field strength can be given as $B=1.6 \delta$\,G. Alternatively, using the fitted turnover (rest-frame) frequency value of, $\nu_\textrm{j}=3.6$\,GHz, one can obtain a much lower magnetic field strength of $B=0.083 \delta$\,G.

Since there is no indication of substantial relativistic boosting in the source, the Doppler factor is expected to have a value below $10$, the above estimated magnetic field strength remains well below the one obtained by \cite{Belladitta_2019}, $B\approx9$\,G. However, the value derived by \cite{Belladitta_2019} characterizes the magnetic field strength at close proximity (fraction of a parsec) to the black hole, while the one estimated from the radio jet is much farther away from the central engine.

Additionally, the above calculation of the magnetic field strength relies on the brightness temperature and size estimations, which may only be limiting values (upper limit on the actual source size, thus lower limit on the brightness temperature) due to the resolution. Therefore, this can also hinder the comparison of the magnetic field strength derived from the X-ray observations and from radio data.

\subsection{Radio power}
We can use the derived spectral index and flux densities to calculate the monochromatic radio powers \citep{hogg2002}: 
\begin{equation}
    P_\nu=4\pi D_\textrm{L}^2 S_\nu (1+z)^{-\alpha-1}
\end{equation}

The obtained radio power values are 
$P_{1.7}=(7.9 \pm 0.8) \times 10^{27}\,\mathrm{W\,Hz}^{-1}$ and $P_{8.5}=(4.6 \pm 1.4) \times 10^{27}\,\mathrm{W\,Hz}^{-1}$. Compared to other high-redshift radio-loud AGN, J0141$-$5427 is among the most powerful ones in the radio regime \citep{Coppejans2016,sotnikova,Mate_APJS}.

\subsection{J0141$-$5427 as a potential VLBI reference source}
\label{subs:refs}

The sky density of known compact bright extragalactic radio sources suitable as VLBI calibrators at declinations below about $-40^\circ$ is significantly lower than at higher declinations \citep[e.g.,][]{icrf3}. This is because most VLBI networks operate on the northern hemisphere. While J0141$-$5427 with its $8.5$-GHz flux density of $\sim 47$\,mJy (Sect.~\ref{sec:res}) is not bright enough for the inclusion in the regular geodetic VLBI observational programmes \citep[e.g.,][]{austral-geod}, it may serve as a phase-reference source for observing weaker nearby targets for high-resolution imaging or relative astrometric positioning. This is especially true at lower frequencies, as indicated by the high rate of fringe-fit solutions found for J0141$-$5427 in our experiment at $1.7$~GHz. So far, VLBI imaging surveys of low-declination southern radio AGN have mainly concentrated on bright sources with at least $\sim 100$\,mJy flux densities \citep[e.g.,][]{sheve1,sheve2,icrf-s1,icrf-s2,tanami1,tanami2}.

\section{Other X-ray weak blazar candidates}

Since J0141$-$5427 is the only known blazar candidate at high redshift with an intense radio but with a very weak X-ray emission, \cite{Belladitta_2019} searched for similar X-ray weak radio-bright blazar candidates in the local Universe using the $5$th edition of the Roma-BZCAT multifrequency catalogue of blazars \citep{roma_bzcat}. They selected flat-spectrum radio sources with flux densities measured at $1.4$\,GHz or $843$\,MHz exceeding $1.5$\,Jy. All these sources have X-ray detections. The authors focused only on sources with $1.4$-GHz radio power similar to that of J0141$-$5427. They found only two objects ($2$\,\% of their sample) with as low X-ray-to-radio luminosity ratio as for J0141$-$5427. 

5BZQ\,J2206$-$1835 is a quasar at redshift $z=0.619$ \citep{z_2206}. It was observed in the prelaunch survey of the VLBI Space Observatory Programme by \citet{vsop1} at $5$\,GHz. It was detected only at the shortest baselines of the Very Long Baseline Array (VLBA). In a $22$\,GHz VLBA survey, \citet{vlba22} did not detect the source. Thus these high-resolution observations did not confirm the blazar nature of 5BZQ\,J2206$-$1835, as they failed to reveal any bright compact radio-emitting feature at mas scale.

5BZQ\,J2038$+$5119, also known as 3C\,418, is a quasar at a redshift of $z=1.686$ \citep{z_2038}. It was observed within the framework of the MOJAVE \citep{Mojave2019} survey at $15$\,GHz. It has a one-sided jet structure with apparent superluminal motion exceeding $6c$ \citep{Mojave2019}. The brightness temperature of the core component is between $4.6 \times 10^{11}$\,K and $5.2 \times 10^{12}$\,K \citep[according to the brightness distribution model of the jet obtained given in ][]{Mojave2019}, thus it exceeds the equipartition limit and implies Doppler boosting. The object was also detected in $\gamma$-rays by the Large Area Telescope onboard the {\it Fermi} satellite \citep{4fgl}.

Thus, the three similarly weak at X-ray radio-loud AGN exhibit very different radio characteristics, forming a heterogeneous group. One of them is a genuine relativistically boosted blazar, another one is not a blazar according to its VLBI observations, and J0141$-$5427 has a modest measured brightness temperature, however, it is compact enough to be detected on intercontinental radio interferometric baselines.

\section{Summary}
\label{sec:sum}

\cite{Belladitta_2019} reported the discovery of a radio-loud AGN at a redshift of $z=5$, which they identified as a possible blazar. Contrary to the expectations, the X-ray emission of this source, J0141$-$5427, is very weak. 

We performed mas-scale resolution radio imaging observations of J0141$-$5427 using the Australian LBA at $1.7$ and $8.5$\,GHz. We detected a single bright, compact feature at both frequencies. This and the flat radio spectrum of the mas-scale feature strengthen its blazar classification. The estimated brightness temperature values clearly indicate the AGN origin of the radio emission.

The relatively low brightness temperature value measured at the rest-frame frequency of $\sim 50$\,GHz is in accordance with the findings of \cite{Cheng2020}. Thus, it still allows for moderate relativistic Doppler boosting that could be directly observable at a lower frequency, in support of the blazar nature of the source. High-resolution VLBI imaging at observed frequencies below $1$\,GHz can sample the assumed turn-over region in the brightness temperature values and provide a Doppler factor for J0141$-$5427. However, such low-frequency ($\lesssim1$\,GHz), high-resolution observations are currently not achievable. 

We investigated the radio properties of two other blazar candidates which have similarly low X-ray-to-radio luminosity ratios as J0141$-$5427. We found that while one of them (J2038$+$5119) clearly shows relativistically boosted radio emission, the other one (J2206$-$1835) is certainly not a blazar.

J0141$-$5427 was detected in X-ray so far in only one observation in $2005$, while remained undetected in $2018$ \citep{Belladitta_2019}. Since blazars are known to show significant variability, a new X-ray observation may provide a better constraint on the high-energy properties of this source. 

%\input{example-content}

% PASA uses footnotes, not endnotes. \endnote in this template will behave like \footnote; and \printendnotes will not output anything.
% \printendnotes

%\bibliography{ref}

%\appendix

%\input{example-appendices}

\begin{acknowledgement}
We thank the referee for his useful feedback that have improved this manuscript.
The Long Baseline Array is part of the Australia Telescope National Facility (\url{https://ror.org/05qajvd42}, accessed 2022.09.10) which is funded by the Australian Government for operation as a National Facility managed by CSIRO. This work was supported by resources provided by the Pawsey Supercomputing Centre with funding from the Australian Government and the Government of Western Australia. The ASKAP radio telescope is part of the Australia Telescope National Facility which is managed by Australia's national science agency, CSIRO. Operation of ASKAP is funded by the Australian Government with support from the National Collaborative Research Infrastructure Strategy. ASKAP uses the resources of the Pawsey Supercomputing Research Centre. Establishment of ASKAP, the Murchison Radio-astronomy Observatory and the Pawsey Supercomputing Research Centre are initiatives of the Australian Government, with support from the Government of Western Australia and the Science and Industry Endowment Fund. We acknowledge the Wajarri Yamatji people as the traditional owners of the Observatory site. This paper includes archived data obtained through the CSIRO ASKAP Science Data Archive, CASDA (https://data.csiro.au). LIG  acknowledges support by the CSIRO Distinguished
Visitor Programme. 
HC acknowledges support from the Hebei Natural Science Foundation of China (Grant No. A2022408002), and the National Natural Science Foundation of China (Grants No. U2031116 and U1731103).
This research was supported by the Australian Research Council Centre of 
Excellence for All Sky Astrophysics in three Dimensions (ASTRO-3D), 
through project number CE170100013. KR acknowledges support from the 
Bundesministerium f\"ur Bildung und Forschung (BMBF) award 05A20WM4.
This research was supported by the Hungarian National Research, Development and Innovation Office (NKFIH), grant number OTKA K134213.
\end{acknowledgement}

\end{document}